\documentclass[aps,prl,reprint,superscriptaddress]{revtex4-1}

\usepackage[latin9]{inputenc}
\usepackage{amsmath}
\usepackage{graphicx}
\usepackage{natbib}
\usepackage{float}

\makeatletter
\begin{document}

\title{Can ultrastrong coupling change ground-state chemical reactions?}

\author{Luis A. Mart\'inez-Mart\'inez}

\affiliation{Department of Chemistry and Biochemistry, University of California
San Diego, La Jolla, California 92093, United States}

\author{Raphael F. Ribeiro}

\affiliation{Department of Chemistry and Biochemistry, University of California
San Diego, La Jolla, California 92093, United States}

\author{Jorge Campos-Gonz\'alez-Angulo}

\affiliation{Department of Chemistry and Biochemistry, University of California
San Diego, La Jolla, California 92093, United States}

\author{Joel Yuen-Zhou}

\affiliation{Department of Chemistry and Biochemistry, University of California
San Diego, La Jolla, California 92093, United States}

\date{\today}
\begin{abstract}
Recent advancements on the fabrication of organic micro- and nanostructures
have permitted the strong collective light-matter coupling regime
to be reached with molecular materials. Pioneering works in this direction
have shown the effects of this regime in the excited state reactivity
of molecular systems and at the same time have opened up the question
of whether it is possible to introduce any modifications in the electronic
ground energy landscape which could affect chemical thermodynamics
and/or kinetics. In this work, we use a model system of many molecules
coupled to a surface-plasmon field to gain insight on the key parameters
which govern the modifications of the ground-state Potential Energy
Surface (PES). Our findings confirm that the energetic changes \textit{per
molecule} are determined by effects which are
essentially on the order of single-molecule light-matter couplings, in contrast with those of the electronically excited
states, for which energetic corrections are of a collective nature.
Still, we reveal some intriguing quantum-coherent effects associated
with pathways of concerted reactions, where two or more molecules
undergo reactions simultaneously, and which can be of relevance in
low-barrier reactions. Finally, we also explore modifications to nonadiabatic
dynamics and conclude that, for our particular model, the presence
of a large number of dark states yields negligible effects. Our study
reveals new possibilities as well as limitations for the emerging
field of polariton chemistry.
\end{abstract}

\pacs{Ultrastrong coupling, ground state, chemical reactivity}

\maketitle

\section{Introduction}

The advent of nano- and microstructures which enable strong confinement
of electromagnetic fields in volumes as small as $1\times10^{-7}\lambda^{3}$\cite{Kim2015},
$\lambda$ being a characteristic optical wavelength, allows for the
possibility of tuning light-matter interactions that can ``dress"
molecular degrees of freedom and give rise to novel molecular functionalities. Several
recent studies have considered the effects of strong coupling (SC)
between confined light and molecular states, and its applications
in exciton harvesting and transport\cite{Gonzalez-Ballestero2015,Feist2015},
charge transfer\cite{Herrera2016}, Bose-Einstein condensation \cite{Andre2006,Gerace2012,Nguyen2015},
Raman \cite{Strashko2016,delPino2015} and photoluminiscence \cite{Herrera2016a,Melnikau2016}
spectroscopy, and quantum computing \cite{DelPino2014,Hartmann2006,Raimond2001},
among many others \cite{Bellesa2004,Laussy2008,Lidzey1999}. Organic
dye molecules are good candidates to explore SC effects
due to their unusually large transition dipole moment \cite{Tischler2005,Hobson2002,Bellessa2004,Salomon2009}.
More recently, it has been experimentally and theoretically shown
that the rates of photochemical processes for molecules placed inside
nanostructures can be substantially modified \cite{Hutchison2012,Herrera2016,Thomas2016,Galego2016,Flick2016}.
The underlying reason for these effects is that the SC energy scale is comparable to that of
vibrational and electronic degrees of freedom,
as well as the coupling between them \cite{Galego2015a}; this energetic
interplay nontrivially alters the resulting energetic spectrum and
dynamics of the molecule-cavity system. It is important to emphasize
that in these examples, SC is the result of a collective coupling between a single photonic
mode and $N \gg 1$ molecules; single-molecule SC coupling is an important frontier of current research \cite{Chikkaraddy2016}, but our emphasis in this work will be on the $N$ molecule case. Since the energy scale of this collective coupling is larger than the molecular and photonic linewidths, the resulting eigenstates of the system have a mixed photon-matter
character. Understanding these so-called \emph{polariton} states is
relevant to develop a physical picture for the emerging energy landscapes
which govern the aforementioned chemical reactivities. More specifically,
Galego and coworkers \cite{Galego2015a} have recently provided a
comprehensive theoretical framework to explain the role of vibronic
coupling and the validity of the Born-Oppenheimer (BO) approximation
in the SC regime, as well as a possible mechanism for changes in photochemical
kinetics afforded by polaritonic systems \cite{Galego2016}; another
theoretical study that focused on control of electron transfer kinetics was given
by Herrera and Spano \cite{Herrera2016}. Using a model of one or two molecules coupled to a single mode in a cavity, Galego and coworkers noticed that some effects on molecular
systems are collective while others are not; similar findings were
reported by Cwik and coworkers using a multimode model and $N$ molecules \cite{Cwik2016}.
While prospects of photochemical control seem promising, it is still
a relatively unexplored question whether ground-state chemical reactivity can be altered
via polaritonic methods, although recently, George and coworkers have
shown a proof of concept of such feasibility using vibrational SC \cite{Thomas2016}.
Along this line, ultrastrong coupling regime (USC) seems to also provide
the conditions to tune the electronic ground-state energy landscape of molecules
and in turn, modify not only photochemistry, but ground-state chemical reactivity. Roughly speaking, this regime
is reached when $\Omega/\hbar\omega_{0}\geq0.1$, $\Omega$
being the (collective) SC of the emitter ensemble to the electromagnetic field and
$\hbar\omega_{0}$ the energy gap of the molecular transition\cite{Moroz}.
Under USC, the ``nonrotating" terms of the light-matter Hamiltonian acquire
relevance and give rise to striking phenomena such as the dynamical
Casimir effect \cite{Wilson,Stassi2013} and Hawking radiation in
condensed matter systems \cite{Stassi2013}. Furthermore, recent experimental
advances have rendered the USC regime feasible in circuit QED \cite{Niemczyk2010},
inorganic semiconductors \cite{Ciuti2005,Todorov2010}, and molecular systems
\cite{Schwartz2011,george2016multiple}, thus prompting us to explore USC effects on ground-state
chemical reactivity.\\
In this article, we address how this reactivity can be influenced
in the USC by studying a reactive model system consisting of an ensemble
of thiacyanine molecules strongly coupled to the plasmonic
field afforded by a metal, where each of the molecules can undergo cis-trans
isomerization by torsional motion. The theoretical model for the photochemistry
of the single thiacyanine molecule has been previously studied in
the context of coherent control \cite{Hoki2009a}. As we will show,
the prospects of controlling ground-state chemical reactivity or nonadiabatic
dynamics involving the ground state are not promising for this particular
model, given that the alterations of the corresponding PES are negligible on a per-molecule basis. However, we
notice the existence of salient quantum-coherent features associated
with concerted reactions that might be worth considering in models
featuring lower kinetic barriers. \\
This article is organized as follows: in the Theoretical Model section,
we describe the polariton system and its quantum mechanical Hamiltonian.
In Methods, we describe the methodology to perform the relevant calculations
and understand the effects of polariton states on the ground-state
PES of the molecular ensemble. In Results and Discussion we describe
our main findings, and finally, in the Conclusions section, we provide
a summary and an outlook of the problem. 

\section{Theoretical model}

\begin{figure}
\includegraphics[scale=0.4]{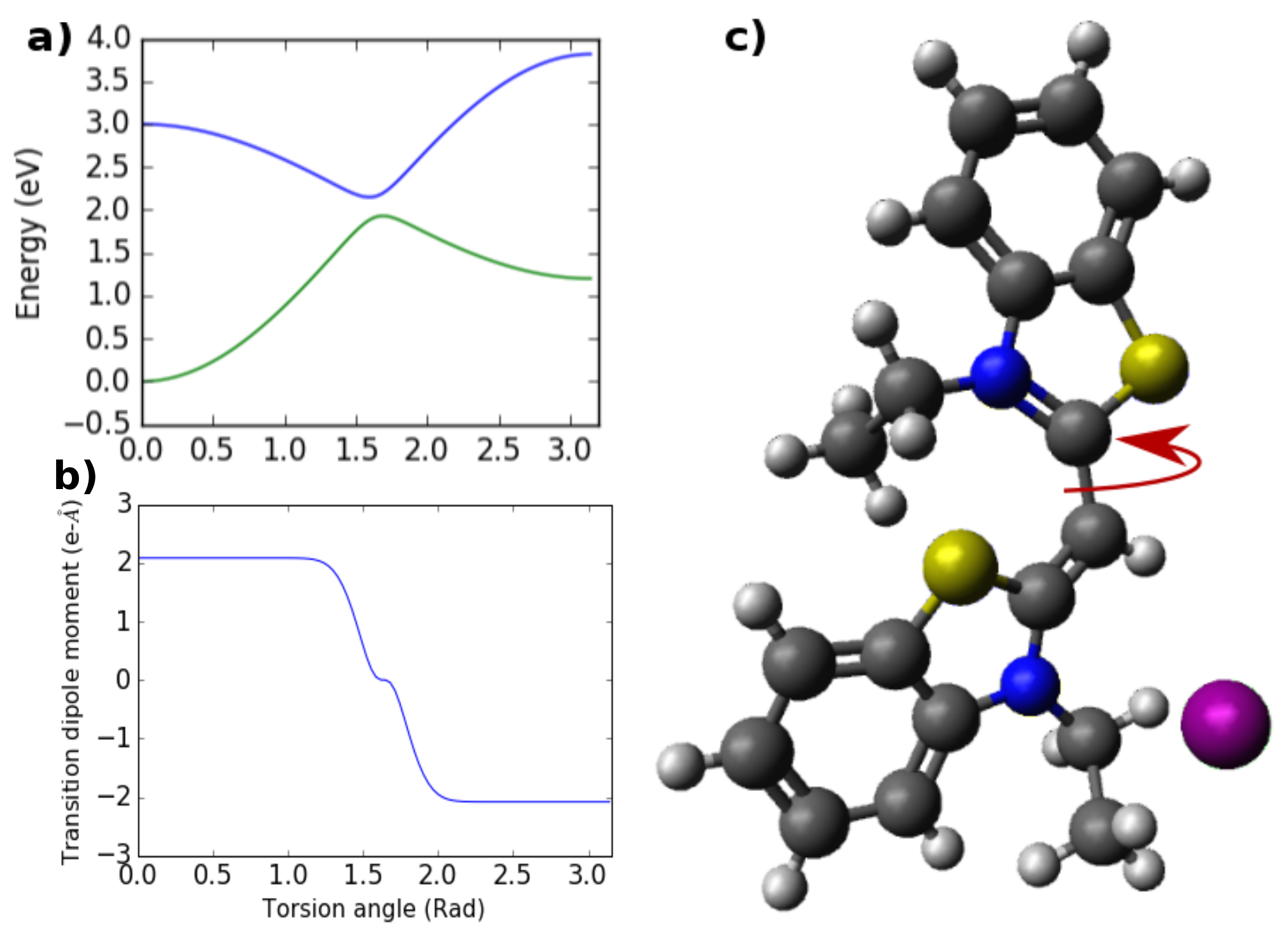} \caption{a) Adiabatic potential energy surfaces (PESs) of the ground and first
excited electronic states of the thyacynine-like model molecule. b)
Transition dipole moment ($\mathbf{\mu}(R)$) of the model molecule
in the adiabatic basis. c) Thiacynine molecule. There exist two geometrical
isomers of the molecule, a cis- and a trans-like configuration. The
cis-trans isomerization of thiacynine-like molecules occurs via a
photo-induced torsion along the bridge which connects the aromatic
rings\label{Adia_PESs}.}
\end{figure}

\begin{figure}
\includegraphics[scale=0.35]{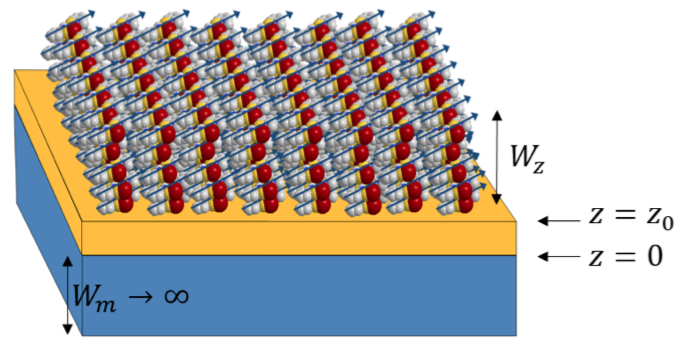} 
\caption{\emph{Plexciton setup.} The model consists of a surface-plasmon (SP)
metal layer whose width $W_{m}$ can be considered infinite in comparison
with the relevant length scales of the structure. The thiacynine molecular
ensemble is separated from the metallic surface by a spacer of width
$z_{0}$; the balls and sticks represent the molecules, while the arrows denote their transition dipole moments. 
The molecular layer has a height $W_{z}$ and is extended along the
$x$ and $y$ planes. \label{fig:plexc_struct}}
\end{figure}

To begin with, we consider a thiacyanine derivative molecule
(Fig. \ref{Adia_PESs}c) and approximate its electronic degrees of
freedom as a quantum mechanical two-level system. To keep the model
tractable, this electronic system is coupled to only one vibrational
degree of freedom $R$, namely, the torsion along the bridge of the
molecule (Fig. \ref{Adia_PESs}c) along which cis-trans isomerization
occurs. The mathematical description of the PES of the ground and
excited states (Fig. \ref{Adia_PESs}a) as well as the transition dipole
moment as a function of the reaction coordinate (Fig. \ref{Adia_PESs}b) have been obtained
from Ref. \cite{Hoki2009a}. The adiabatic representation of the electronic
states is given by,
\begin{equation}
\begin{split}|g(R)\rangle= & \cos\left(\theta(R)/2\right)|\text{trans}\rangle+\sin\left(\theta(R)/2\right)|\text{cis}\rangle\\
|e(R)\rangle= & -\sin\left(\theta(R)/2\right)|\text{trans}\rangle+\cos\left(\theta(R)/2\right)|\text{cis}\rangle
\end{split}
\label{adiabatic_rep}
\end{equation}
where $|e(R)\rangle$ and $|g(R)\rangle$ are the $R$-dependent adiabatic excited
and ground state respectively. $|\text{trans}\rangle$ and $|\text{cis}\rangle$
are the ($R$-independent) crude diabatic electronic states that describe
the localized chemical character of each of the isomers. The ground-state
PES has a predominant trans (cis) character to the left (right) of
the barrier ($\theta(0)=0$, $\theta(\pi)=\pi$) in Fig. \ref{Adia_PESs}a.\\
Our USC model consists of a setup where an orthorhombic ensemble of
thyacyanine molecules is placed on top of a thin spacer which, in
turn, is on top of a metallic surface that hosts surface plasmons
(SPs) \cite{Yuen-Zhou2015} (see Fig. \ref{fig:plexc_struct}). The
coupling between molecular electronic transitions and plasmons in the metal give
rise to polaritons that are often called plexcitons \cite{Gonzalez-Tudela2013,Yuen-Zhou2015}.
The ensemble is comprised of $N_{z}$ single-molecule layers. The location
of each molecule can be defined by the Cartesian coordinates $\mathbf{n}+(0,0,z_{s})$
where $\mathbf{n}=(\Delta_{x}n_{x},\Delta_{y}n_{y},0)$ and $z_{s}=z_{0}+\Delta_{z}s$
for the \textit{s}-th layer. Here, the spacing between molecules along
the \textit{i}-th direction is denoted by $\Delta_{i}$, and $z_{0}$
is the width of the spacer (see Fig. \ref{fig:plexc_struct}). We
chose a SP electromagnetic environment because its evanescent intensity
decreases fast enough with momentum $\mathbf{k}$ (giving rise to
vanishing light-matter coupling for large $|\mathbf{k}|$), resulting
on a convergent Lamb-shift of the molecular ground-state. As shall
be explained below, this circumvents technical complications of
introducing renormalization cutoffs, as would be needed for a dielectric
microcavity \cite{Cwik2016}. The Hamiltonian of the plexciton setup
is given by $H=H_{el}+T_{nuc}$, where $T_{nuc}=\sum_{i}\frac{\mathbf{P}_{i}^{2}}{2M_{i}}$
is the nuclear kinetic energy operator and\begin{widetext}
\begin{equation}
\begin{split}H_{el}(\mathbf{R})= & \sum_{\mathbf{k}}\hbar\omega_{\mathbf{k}}a_{\mathbf{k}}^{\dagger}a_{\mathbf{k}}+\sum_{\mathbf{n},s}\left(\hbar\omega_{e}(R_{\mathbf{n},s})-\hbar\omega_{g}(R_{\mathbf{n},s})\right)b_{\mathbf{n},s}^{\dagger}(R_{\mathbf{n},s})b_{\mathbf{n},s}(R_{\mathbf{n},s})\\
 & +\sum_{\mathbf{k}}\sum_{\mathbf{n},s}g_{\mathbf{k}}^{\mathbf{n},s}(R_{\mathbf{n},s})\left(a_{\mathbf{k}}^{\dagger}b_{\mathbf{n},s}(R_{\mathbf{n},s})+a_{\mathbf{k}}b_{\mathbf{n},s}^{\dagger}(R_{\mathbf{n},s})+a_{\mathbf{k}}b_{\mathbf{n},s}(R_{\mathbf{n},s})+a_{\mathbf{k}}^{\dagger}b_{\mathbf{n},s}^{\dagger}(R_{\mathbf{n},s})\right)\\
 & +\sum_{\mathbf{n},s}\hbar\omega_{g}(R_{\mathbf{n},s}),
\end{split}
\label{Dicke}
\end{equation}
\end{widetext}corresponds to the Dicke Hamiltonian \cite{garraway2011dicke}.
Here $a_{\mathbf{k}}^{\dagger}$ ($a_{\mathbf{k}}$) is the creation
(annihilation) operator for the SP mode with in-plane momentum
$\mathbf{k}$ which satisfies $[a_{\mathbf{k}},a_{\mathbf{k}'}^{\dagger}]=\delta_{\mathbf{k},\mathbf{k}'}$,
and $\mathbf{R}=\{R_{\mathbf{n}s}\}$ is an $N$-dimensional vector
that describes the vibrational coordinates of the $N=N_{x}N_{y}N_{z}$
molecules of the ensemble, where $N_{i}$ is the number of molecules
along each ensemble axis. $\hbar\omega_{g}(R_{\mathbf{n},s})$ accounts
for the ground-state energy of the molecule whose location in the
ensemble is defined by $\mathbf{n}$ and $s$. We introduce the (adiabatic
$R$-dependent) exciton operator $b_{\mathbf{n},s}^{\dagger}(R_{\mathbf{n},s})\left(b_{\mathbf{n},s}(R_{\mathbf{n},s})\right)$
to label the creation (annihilation) of a Frenkel exciton (electronic
excitation) with an energy gap $\hbar\omega_{e}(R_{\mathbf{n},s})-\hbar\omega_{g}(R_{\mathbf{n},s})$
on the molecule located at $\mathbf{n}+z_{s}\mathbf{\hat{z}}$. The
coefficients $\hbar\omega_{\mathbf{k}}$ and $g_{\mathbf{k}}^{\mathbf{n},s}(R_{\mathbf{n},s})$
stand for the energy of a SP with in-plane momentum $\mathbf{k}$ and the coupling
of the molecule located at $\mathbf{n}+z_{s}\hat{\mathbf{z}}$ with
the latter, respectively. The dipolar SP-matter interaction is described by $g_{\mathbf{k}}^{\mathbf{n},s}(R_{\mathbf{n},s})=h_{\mathbf{k}}(R_{\mathbf{n},s})f_{\mathbf{k}}(z_{s})$,
where $h_{\mathbf{k}}(R_{\mathbf{n},s})=-\mathbf{\boldsymbol{\mu}}_{\mathbf{n},s}(R_{\mathbf{n},s})\cdot\mathbf{E}_{\mathbf{k}}(\mathbf{n})$
is the projection of the molecular transition dipole $\mathbf{\boldsymbol{\mu}}_{\mathbf{n},s}(R_{\mathbf{n},s})$
onto the in-plane component of the SP electric field $\mathbf{E}_{\mathbf{k}}(\mathbf{n})$
and $f_{\mathbf{k}}(z_{s})=e^{-\alpha_{\mathbf{k}}z_{s}}$
is the evanescent field profile along the $z$ direction, with $\alpha_{\mathbf{k}}$
being the decay constant in the molecular region ($z>0$). The quantized
plasmonic field $\hat{\mathbf{E}}_{\mathbf{k}}f_{\mathbf{k}}(\mathbf{z}_{s})$
has been discussed in previous works \cite{Novotny,Gonzalez-Tudela2013,Torma2015,Yuen-Zhou2015}
and reads $\hat{\mathbf{E}}_{\mathbf{k}}(\mathbf{n})f_{\mathbf{k}}(\mathbf{z}_{s})=\sqrt{\frac{\hbar\omega_{\mathbf{k}}}{2\epsilon_{0}SL_{\mathbf{k}}}}a_{\mathbf{k}}\hat{\mathbf{\chi}}_{\mathbf{k}}e^{i\mathbf{k}\cdot\mathbf{n}}e^{-\alpha_{\mathbf{k}}z}+h.c.$,
where $\epsilon_{0}$ is the free-space permittivity, $S$ is the
coherence area of the plexciton setup, $L_{\mathbf{k}}$ is the quantization
length, and $\hat{\mathbf{\chi}}_{\mathbf{k}}=\hat{\mathbf{k}}+i\frac{|\mathbf{k}|}{\alpha_{\mathbf{k}}}\hat{\mathbf{z}}$
is the polarization. Note that the parametric dependence of the exciton
operators on $R_{\mathbf{n},s}$ yield residual non-adiabatic processes
induced by nuclear kinetic energy that may be relevant to the isomerization
in question. We also highlight the fact that Eq. (\ref{Dicke}) includes
both rotating (``energy conserving'') terms ($a_{\mathbf{k}}^{\dagger}b_{\mathbf{n},s}$
and $a_{\mathbf{k}}b_{\mathbf{n},s}^{\dagger}$) where a photon creation
(annihilation) involves the concomitant annhilation (creation) of an
exciton; and counterrotating (``non-energy conserving'') terms ($a_{\mathbf{k}}b_{\mathbf{n},s}$
and $a_{\mathbf{k}}^{\dagger}b_{\mathbf{n},s}^{\dagger}$) where there
is a simultaneous annhilation (creation) of photon and exciton. These
latter terms are ignored in the widely used Rotating Wave Approximation
(RWA)\cite{scully1999quantum}, where light-matter coupling is weak compared to the transition energy. Since we are interested in the USC, we shall keep them throughout.
\vspace{20mm}
\section{Methods}
\begin{figure}
\includegraphics[scale=0.34]{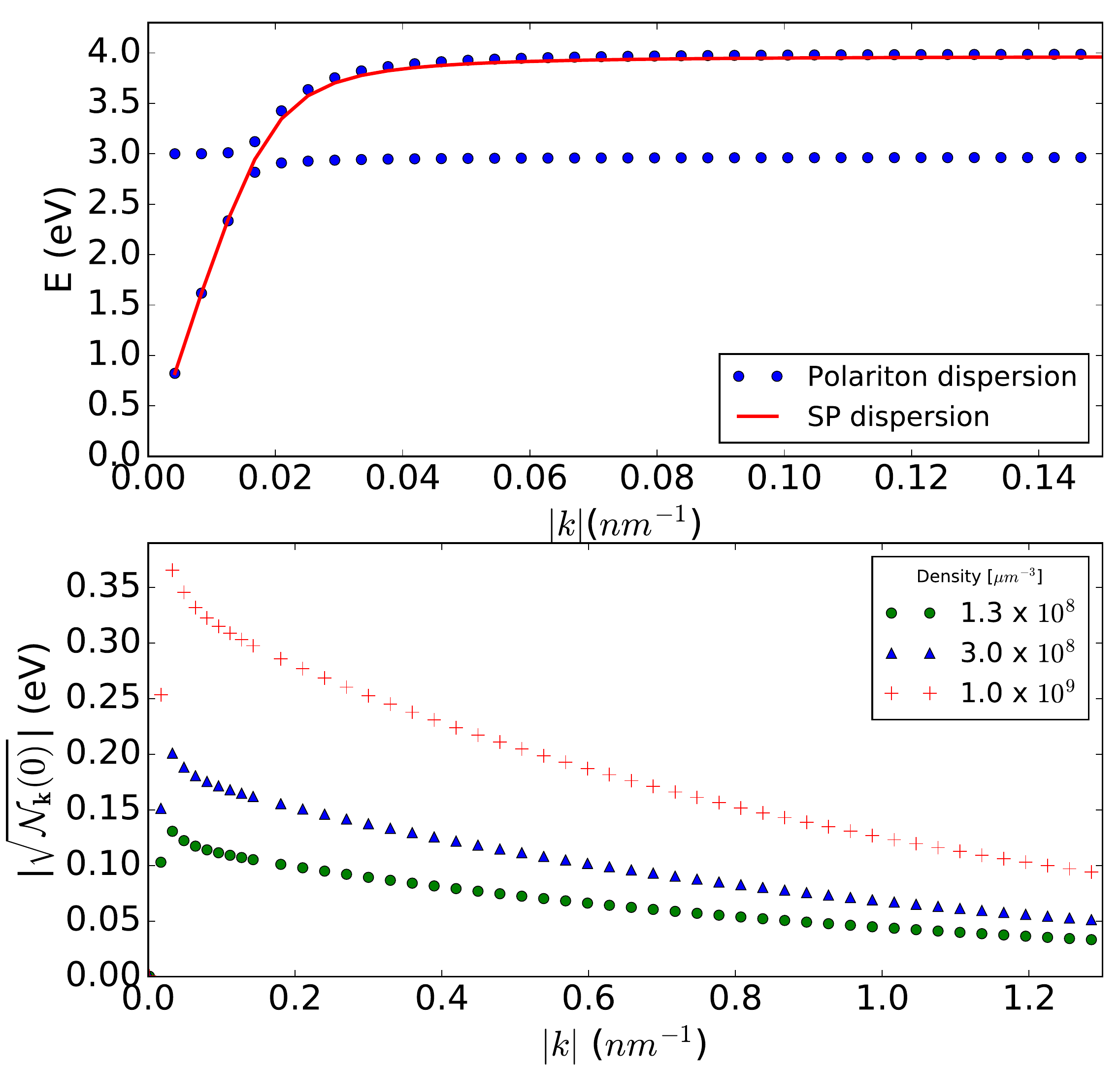}\caption{$\textit{Upper}$:
Polariton dispersion that results from the interaction of a molecular
ensemble with the plasmonic field; we chose $\rho=1.0\times10^{9}\mu m^{-3}$. $\textit{Lower}$: Collective SP-exciton coupling at equilibrium geometry
$\sqrt{\mathcal{N}_{\mathbf{k}}(0)}$ as a function of $|\mathbf{k}|$,
assuming $\mathbf{\mu}(R=0)$ and $\mathbf{k}$ are parallel to the
$x$ axis. We consider a slab with $W_{z}=120$ nm and compute couplings
as a function of varying molecular densities $\rho$. The range of
the resulting couplings is well above the plasmonic linewidth of the
order of 10 $meV$ \cite{Gonzalez-Tudela2013}, indicating the polaritonic
onset of strong and ultrastrong light-matter coupling.
\label{Collective-coupling}}
\end{figure}

For simplicity, we assume that all the transition
dipoles are equivalent and aligned along $x$, $\mathbf{\boldsymbol{\mu}}_{\mathbf{n},s}(R_{\mathbf{n},s})=\boldsymbol{\mu}(R_{\mathbf{n},s})=\mu(R_{\mathbf{n},s})\hat{\mathbf{x}}$;
a departure of this perfect crystal condition does not affect the
conclusions of this article. Furthermore, it is convenient to first
restrict ourselves to the cases where all nuclei are fixed at the
same configuration ($\mathbf{R}=\tilde{\mathbf{R}}$, which denotes
$R_{\mathbf{n},s}=R$ for all $\mathbf{n}$ and $s$), so that we
can take advantage of the underlying translational symmetry to introduce
a delocalized exciton basis where the in-plane momentum $\mathbf{k}$ is a
good quantum number. The creation operator of this delocalized state
is defined by $b_{\mathbf{k}}^{\dagger}(R)=\frac{1}{\sqrt{\mathcal{N}_{\mathbf{k}}(R)}}\sum_{\mathbf{n}}\sum_{s}f_{\mathbf{k}}(z_{s})h_{\mathbf{k}}(R)b_{\mathbf{n},s}^{\dagger}(R)$,
and the normalization squared is given by $\mathcal{N}_{\mathbf{k}}(R)=\sum_{\mathbf{n}}\sum_{s}|h_{\mathbf{k}}(R)|^{2}|f_{\mathbf{k}}(z_{s})|^{2}$
which, in the continuum limit, can be seen to be proportional to $\rho$, the number density of the molecular ensemble. In this
collective basis, the previously introduced $H_{el}(\mathbf{R})$ reads 
\begin{widetext}
\begin{equation}
\begin{split}
H_{el}(\tilde{\mathbf{R}}) & =\sum_{\mathbf{k}}\hbar\Delta(R)b_{\mathbf{k}}^{\dagger}(R)b_{\mathbf{k}}(R)+\sum_{\mathbf{k}}\hbar\omega_{\mathbf{k}}a_{\mathbf{k}}^{\dagger}a_{\mathbf{k}}\\
 & +\sum_{\mathbf{k}}\sqrt{\mathcal{N}_{\mathbf{k}}(R)}\left(a_{\mathbf{k}}^{\dagger}b_{\mathbf{k}}(R)+a_{\mathbf{k}}b_{\mathbf{k}}^{\dagger}(R)+a_{\mathbf{k}}b_{-\mathbf{k}}(R)+a_{\mathbf{k}}^{\dagger}b_{-\mathbf{k}}^{\dagger}(R)\right)+\sum_{\mathbf{k}}H_{\mathrm{dark},\mathbf{k}}(R)+\sum_{\mathbf{k}}H_{\mathrm{unklapp},\mathbf{k}}(R)+N\hbar\omega_{g}(R)\\
 & =\sum_{\mathbf{k}}H_{\mathbf{k}}(R)+\sum_{\mathbf{k}}H_{\mathrm{dark},\mathbf{k}}(R)+\sum_{\mathbf{k}}H_{\mathrm{unklapp},\mathbf{k}}(R)+N\hbar\omega_{g}(R),
\end{split}
\label{collective}
\end{equation}
\end{widetext}where $\Delta(R)=\omega_{e}(R)-\omega_{g}(R)$ is the exciton transition
frequency. 
\begin{equation}\label{dark_states}
H_{\mathrm{dark},\mathbf{k}}(R)	=\hbar\Delta(R)\mathbf{P}_{\mathrm{dark},\mathbf{k}}(R)
\end{equation}
accounts for the energy of the ($N_{z}-1$)-degenerate exciton states with in-plane momentum $\mathbf{k}$ that do not couple to SPs, and are usually known as \textit{dark} \textit{states}. The latter are orthogonal to the bright exciton $b_{\mathbf{k}}^{\dagger}(R)|G_{m}(\tilde{\mathbf{R}})\rangle$ that couples to the SP field, where $|G_{m}(\tilde{\mathbf{R}})\rangle$ is the bare molecular ground-state ($b_{\mathbf{k}}(R)|G_{m}(\tilde{\mathbf{R}})\rangle=0$).
 More specifically, $\mathbf{P}_{\mathrm{dark},\mathbf{k}}(R)=\mathbf{I}_{\mathrm{exc},\mathbf{k}}(R)-b_{\mathbf{k}}^{\dagger}(R)b_{\mathbf{k}}(R)$ is a projector operator onto the $\mathbf{k}$-th dark-state subspace, with $\mathbf{I}_{\mathrm{exc}}(R)=\sum_{\mathbf{n},s}b_{\mathbf{n},s}^{\dagger}(R)b_{\mathbf{n},s}(R)=\sum_{\mathbf{k},s}b_{\mathbf{k},s}^{\dagger}(R)b_{\mathbf{k},s}(R)=\sum_{\mathbf{k}}\mathbf{I}_{\mathrm{exc},\mathbf{k}}(R)$ being the identity on the exciton space, and $b_{\mathbf{k},s}^{\dagger}(R)=\frac{1}{\sqrt{N_{x}N_{y}}}\sum_{\mathbf{n}}e^{-i\mathbf{k}\cdot\mathbf{n}}b_{\mathbf{n},s}^{\dagger}(R)$. Finally,
\begin{equation}
\begin{split}
H_{\mathrm{unklapp},\mathbf{k}}(R)=&\sum_{\mathbf{q}=(\frac{2\pi q_{x}}{\Delta_{x}},\frac{2\pi q_{y}}{\Delta_{y}})}\sqrt{\mathcal{N}_{\mathbf{k}}(R)}\Big(a_{\mathbf{k+q}}^{\dagger}b_{\mathbf{k}}(R)\\
&+a_{\mathbf{k}+q}^{\dagger}b_{-\mathbf{k}}^{\dagger}(R)+h.c.\Big)
\end{split}
\end{equation}
stands for the coupling of excitons with momentum $\mathbf{k}$
to SP modes with momentum beyond the first excitonic Brillouin zone.
$H_{\mathrm{unklapp},\mathbf{k}}(R)$ is usually ignored given the large off-resonance
between the SP energy and the exciton states; however, since this
work pertains off-resonant effects, we considered it to acquire converged
quantities in the calculations explained below. We also note that
the normalization constant $\sqrt{\mathcal{N}_{\mathbf{k}}(R)}$ in
Eq. \ref{collective} is precisely the collective SP-exciton coupling.
As mentioned in the introduction, the condition $\sqrt{\mathcal{N_{\mathbf{k}}}(R)}/\hbar\Delta(R)>0.1$
is often used to define the onset of USC \cite{Moroz}, and it is fulfilled with the maximal density considered in our model (see Fig. \ref{Collective-coupling}) taking into account that the largest $\hbar \Delta(R)$ is 3 eV (See Fig. \ref{Adia_PESs}a). We note, as will be evident later, that our main results do not vary significantly by considering ratios $\sqrt{\mathcal{N_{\mathbf{k}}}(R)}/\hbar\Delta(R)$ below the aforementioned threshold. 

A Bogoliubov transformation \cite{Ciuti2005} permits the diagonalization
of the Bloch Hamiltonian $H_{\mathbf{k}}$ in Eq. \ref{collective} by
introducing the polariton quasiparticle operators 
\begin{equation}
\xi_{\mathbf{k}}^{j}(R)=\alpha_{\mathbf{k}}^{j}a_{\mathbf{k}}+\beta_{\mathbf{k}}^{j}b_{\mathbf{k}}(R)+\gamma_{\mathbf{k}}^{j}a_{-\mathbf{k}}^{\dagger}+\delta_{\mathbf{k}}^{j}b_{-\mathbf{k}}^{\dagger}(R),\label{eq:bogoliubov_quasi}
\end{equation}
where $j=U,L$ and $U$ ($L$) stands for the upper (lower) Bogoliubov polariton
state. Notice that this canonical transformation is valid for a sufficiently
large number of molecules $N$, where the collective exciton operators
$b_{\mathbf{k}}(R)$, $b_{\mathbf{k}}^{\dagger}(R)$ are well approximated
by bosonic operators \cite{Tassone1999}.\\
The bare molecular ground-state with no photons in the absence of light-matter coupling $|G_{m}(\tilde{\mathbf{R}});0\rangle$, ($a_{\mathbf{k}}|G_{m}(\tilde{\mathbf{R}});0\rangle=b_{\mathbf{k}}(R)|G_{m}(\tilde{\mathbf{R}});0\rangle=0$
for all $\mathbf{k}$) has a total extensive energy with molecular contributions only $\langle G_{m}(\tilde{\mathbf{R}});0|H_{el}(\tilde{\mathbf{R}})|G_{m}(\tilde{\mathbf{R}});0\rangle=N\hbar\omega_{g}(R)$.
Upon inclusion of the counterrotating terms, the ground-state
becomes the dressed Bogoliubov vacuum $|G(\tilde{\mathbf{R}})\rangle_{d}$,
characterized by $\xi_{\mathbf{k}}^{j}(R)|G(\tilde{\mathbf{R}})\rangle_{d}=0$
for all $\mathbf{k}$ and $j$, with total energy $_{d}\langle G(\tilde{\mathbf{R}})|H_{el}(\tilde{\mathbf{R}})|G(\tilde{\mathbf{R}})\rangle_{d}=E_{0}(\tilde{\mathbf{R}})$,
where the zero-point energy is given by
\begin{equation}
\begin{split}
E_{0}(\tilde{\mathbf{R}})=&N\hbar\omega_{g}(R)\\
&+\frac{1}{2}\sum_{\mathbf{k}}\left(\sum_{j=U,L}\hbar\omega_{j,\mathbf{k}}(R)-\hbar\omega_{\mathbf{k}}-\hbar\Delta(R)\right),
\end{split}\label{shift_vacuum}
\end{equation}
$\{\hbar\omega_{j,\mathbf{k}}(R)\}$ being the eigenvalues of the
Bogoliubov polariton branches given by
\begin{equation}
\omega_{_{_{L}^{U}},\mathbf{k}}(R)=\sqrt{\frac{(\Delta(R))^{2}+\omega_{\mathbf{k}}^{2}\pm\sqrt{B(R)^{2}+16\mathcal{N}_{\mathbf{k}}^{2}(R)\Delta(R)\omega_{\mathbf{k}}}}{2}},\label{eq:bogo_disp}
\end{equation}
where we have introduced $B(R)=\omega_{\mathbf{k}}^{2}-\Delta(R)^{2}$. A hallmark of the SC and USC regimes is the anticrossing splitting
of the polariton energies at the $\mathbf{k}$ value where the bare
excitations are in resonance, $\Delta(R)=\omega_{\mathbf{k}}$ \cite{Torma2015} (see Fig. \ref{Collective-coupling}).
The sum in Eq. \ref{shift_vacuum} accounts for the energy shift from
the bare molecular energy $N\hbar\omega_{g}(R)$ due to interaction with the
infinite number of SP modes in the setup. Using Eq. (\ref{eq:bogo_disp}),
it is illustrative to check that this shift vanishes identically when
the non-RWA terms are ignored. 

It is worth describing some of the physical aspects of the Bogoliubov
ground-state $|G(\tilde{\mathbf{R}})\rangle_{d}$. With the numerically
computed wavefunctions, we can use the inverse transformation of Eq.
\ref{eq:bogoliubov_quasi} to explicitly evaluate its SP and exciton
populations \cite{Ciuti2005},

\begin{subequations}\label{eq:populations}
\begin{align}
n_{\mathbf{k}}^{SP}={}_{d}\langle G(\tilde{\mathbf{R}})|a_{\mathbf{k}}^{\dagger}a_{\mathbf{k}}|G(\tilde{\mathbf{R}})\rangle_{d} & =\sum_{j}|\gamma_{\mathbf{k}}^{j}|^{2},\label{eq:pop_SP}\\
n_{\mathbf{k}}^{exc}={}_{d}\langle G(\tilde{\mathbf{R}})|b_{\mathbf{k}}^{\dagger}b_{\mathbf{k}}|G(\tilde{\mathbf{R}})\rangle_{d} & =\sum_{j}|\delta_{\mathbf{k}}^{j}|^{2},\label{eq:pop_exc}
\end{align}
\end{subequations}which give rise to humble $O(10^{-3})$ values per
mode $\mathbf{k}$, considering a molecular ensemble with $\rho=3\times 10^{8} \mu m^{-3}$ and $W_{z}=$120 nm; this calculation is carried out using $N=8\times10^{7}$, although results are largely insensitive to this parameter as long as it is sufficiently large to capture the thermodynamic limit. The consequences of the dressing partially accounted for by Eq. (\ref{eq:populations}) (partially since there are also correlations of the form $_{d}\langle G(\tilde{\mathbf{R}})|b_{\mathbf{k}}a_{-\mathbf{k}}|G(\tilde{\mathbf{R}})\rangle_{d}$) are manifested as energetic effects on $|G_{m}(\tilde{\mathbf{R}});0\rangle$: $E_{0}(\tilde{\mathbf{R}})-N\hbar\omega_{g}(R)$
can be interpreted as the energy stored in $|G(\tilde{\mathbf{R}})\rangle_{d}$
as a result of dressing; it is an extensive quantity
of the ensemble, but becomes negligible when considering a per-molecule
stabilization. For instance, in molecular ensembles with the aforementioned parameters we find $E_{0}(\tilde{\mathbf{0}})-N\hbar\omega_{g}(0)=O(10^{2})$
eV, which implies a $O(10^{-5})$ eV value per molecule; our calculations
show that this intensive quantity is largely insensitive to total number
of molecules. This observation raises the following questions: to what extent does photonic
dressing would impact ground-state chemical reactivity? What are
the relevant energy scales that dictate this impact?
With these questions in mind, we aim to study the polaritonic effects on ground-state single-molecule isomerization
events. To do so, we map out the PES cross section where we set one
``free\char`\"{} molecule to undergo isomerization while fixing the
rest at $R_{\mathbf{n},s}=0$. A similar strategy has been used before in \cite{Galego2016}.
This cross section, described by $E_{0}(R_{\mathbf{n}_{0},0},0,\cdots,0)\equiv E_{0}(R_{\mathbf{n}_{0},0},\mathbf{\tilde{0}}')$
($R_{\mathbf{n}_{0},0}$ being the coordinate of the unconstrained molecule),
should give us an approximate understanding of reactivity starting
from thermal equilibrium conditions, since the molecular configuration $\tilde{\mathbf{R}}=\tilde{\mathbf{0}}$ still corresponds
to the global minimum of the modified ground-state PES, as will be
argued later. By allowing one molecule to move differently
than the rest, we weakly break translational symmetry. Rather than
numerically implementing another Bogoliubov transformation, we can, to a very good approximation, account for this motion by treating the isomerization
of the free molecule as a perturbation on $H_{el}(\tilde{\boldsymbol{0}})$.
More precisely, we write $H_{el}(R_{\mathbf{n}_{0},0},\mathbf{\tilde{0}}')|G(R_{\mathbf{n}_{0},0},\mathbf{\tilde{0}}')\rangle_{d}=E_{0}(R_{\mathbf{n}_{0},0},\mathbf{\tilde{0}}')|G(R_{\mathbf{n}_{0},0},\mathbf{\tilde{0}}')\rangle_{d}$,
where $H_{el}(R_{\mathbf{n}_{0},0},\mathbf{\tilde{0}}')$ is the sum
of a translationally invariant piece $H_{el}(\tilde{\mathbf{0}})$
plus a perturbation due to the free molecule,
\begin{equation}
H_{el}(R_{\mathbf{n}_{0},0},\mathbf{\tilde{0}}')=H_{el}(\tilde{\mathbf{0}})+V(R_{\mathbf{n}_{0},0}).\label{def_pert}
\end{equation}
The perturbation is explicitly given by 
\begin{widetext}
\begin{equation}
\begin{split}V(R_{\mathbf{n}_{0},0}) & =H_{el}(R_{\mathbf{n}_{0},0},\mathbf{\tilde{0}}')-H_{el}(\tilde{\mathbf{0}})\\
 & =\hbar\Delta(R_{\mathbf{n}_{0},0})b_{\mathbf{n}_{0},0}^{\dagger}(R_{\mathbf{n}_{0},0})b_{\mathbf{n}_{0},0}(R_{\mathbf{n}_{0},0})-\hbar\Delta(0)b_{\mathbf{n}_{0},0}^{\dagger}(0)b_{\mathbf{n}_{0},0}(0)\\
 & +\sum_{\mathbf{k}}\Bigg\{ g_{\mathbf{k}}^{\mathbf{n}_{0},0}(R_{\mathbf{n}_{0},0})\Bigg[b_{\mathbf{n}_{0},0}(R_{\mathbf{n}_{0},0})+b_{\mathbf{n}_{0},0}^{\dagger}(R_{\mathbf{n}_{0},0})\Bigg]-g_{\mathbf{k}}^{\mathbf{n}_{0},0}(0)\Bigg[b_{\mathbf{n}_{0},0}(0)+b_{\mathbf{n}_{0},0}^{\dagger}(0)\Bigg]\Bigg\}\Bigg[a_{\mathbf{k}}+a_{\mathbf{k}}^{\dagger}\Bigg]\\
 & +\hbar\omega_{g}(R_{\mathbf{n}_{0},0})-\hbar\omega_{g}(0).
\end{split}
\label{perturbation}
\end{equation}
\end{widetext}
Notice that we have chosen the free molecule to be located at an arbitrary
in-plane location $\mathbf{n}_{0}$ and at the very bottom of the
slab at $s=0$, where light-matter coupling is strongest as a result
of the evanescent field profile along the $z$ direction. We write
an expansion of the PES cross section as $E_{0}(R_{\mathbf{n}_{0},0},\mathbf{\tilde{0}}')=\sum_{q=0}^{\infty}E_{0}^{(q)}(R_{\mathbf{n}_{0},0},\mathbf{\tilde{0}}')$,
where $q$ labels the $O(V^{q})$ perturbation correction. The zeroth
order term is the Bogoliubov vacuum energy associated to every molecule
being at the equilibrium geometry $E_{0}^{(0)}(R_{\mathbf{n}_{0},0},\mathbf{\tilde{0}}')=E_{0}(\tilde{\mathbf{0}})$
as in Eq. (\ref{shift_vacuum}). The $O(V)$ correction corresponds
to $\hbar\omega_{g}(R_{\mathbf{n}_{0},0})-\hbar\omega_{g}(0)$,
merely describing the PES of the isomerization of the bare molecule
in the absence of coupling to the SP field. The contribution of the
SP field on the PES cross-section of interest appears at $O(V^{2})$,
and it is given by
\begin{align}
E^{(2)}(R_{\mathbf{n}_{0},0},\mathbf{\tilde{0}}')\approx\sum_{_{i,j=UP,LP}^{\mathbf{k}_{1}\leq\mathbf{k}_{2}}}\frac{|\langle\mathbf{k}_{1},i;\mathbf{k}_{2},j|V(R_{\mathbf{n}_{0},0})|G(\tilde{\mathbf{0}})\rangle_{d}|^{2}}{E_{0}(\tilde{\mathbf{0}})-E_{\mathbf{k}_{1},\mathbf{k}_{2},i,j}^{(0)}},\label{second_order}
\end{align}
where $|\mathbf{k}_{1},i;\mathbf{k}_{2},j\rangle\equiv\xi_{\mathbf{k}_{1}}^{i\dagger}(0)\xi_{\mathbf{k}_{2}}^{\dagger j}(0)|G(\tilde{\mathbf{0}})\rangle_{d}$
and $E_{\mathbf{k}_{1},\mathbf{k}_{2},i,j}^{(0)}=\hbar(\omega_{i,\mathbf{k}_{1}}(0)+\omega_{j,\mathbf{k}_{2}}(0))$.
As shown in the Appendix, the approximation in Eq. (\ref{second_order})
consists of ignoring couplings between $|G(\tilde{\mathbf{0}})\rangle_{d}$
and states with three and four Bogoliubov polariton excitations, since
their associated matrix elements become negligible in the thermodynamic
limit compared to their double excitation counterparts. The remaining
matrix elements can be calculated by expressing the operators $a_{\mathbf{k}},\,a_{\mathbf{k}}^{\dagger},\,b_{\mathbf{n}_{0},0}(R_{\mathbf{n}_{0},0}),\,b_{\mathbf{n}_{0},0}^{\dagger}(R_{\mathbf{n}_{0},0})$
in Eq. (\ref{perturbation}) in terms of the Bogoliubov operators
$\xi_{\mathbf{k}}^{j}(0),\,\xi_{\mathbf{k}}^{\dagger j}(0)$ (see
Eq. (\ref{eq:bogoliubov_quasi})), leading to
\begin{widetext}
\begin{equation}
\begin{split}\langle\mathbf{k}_{1},i;\mathbf{k}_{2},j|V(R_{\mathbf{n}_{0},0})|G(\tilde{\mathbf{0}})\rangle_{d} & =F^{\mathbf{k}_{2}}(R_{\mathbf{n}_{0},0})D_{\mathbf{k}_{1}}\left(-\delta_{-\mathbf{k}_{1}}^{i}\alpha_{\mathbf{k}_{2}}^{j}+\delta_{-\mathbf{k}_{1}}^{i}\gamma_{-\mathbf{k}_{2}}^{j}-\beta_{\mathbf{k}_{1}}^{i}\gamma_{-\mathbf{k}_{2}}^{j}+\beta_{\mathbf{k}_{1}}^{i}\alpha_{\mathbf{k}_{2}}^{j}\right)\\
 & +F^{\mathbf{k}_{1}}(R_{\mathbf{n}_{0},0})D_{\mathbf{k}_{2}}\left(-\delta_{-\mathbf{k}_{2}}^{j}\alpha_{\mathbf{k}_{1}}^{i}+\delta_{-\mathbf{k}_{2}}^{j}\gamma_{-\mathbf{k}_{1}}^{i}-\beta_{\mathbf{k}_{2}}^{j}\gamma_{-\mathbf{k}_{1}}^{i}+\beta_{\mathbf{k}_{2}}^{j}\alpha_{\mathbf{k}_{1}}^{i}\right),
\end{split}
\label{coupling}
\end{equation}
\end{widetext}where $F^{\mathbf{k}}(R)=\cos(\theta(R))g_{\mathbf{k}}^{\mathbf{n}_{0},0}(R)-\cos(\theta(0))g_{\mathbf{k}}^{\mathbf{n}_{0},0}(0)$ depends on
the mixing angle that describes the change of character of $b_{\mathbf{n}_{0},0}^{\dagger}(R)$
as a function of $R$ (see Equation (\ref{adiabatic_rep})); it emerges
as a consequence of coupling molecular states at different configurations.
$D_{\mathbf{k}}=\langle G_{m}(\tilde{\mathbf{R}});0\rangle|b_{\mathbf{n}_{0},0}(0)b_{\mathbf{k}}^{\dagger}(0)|G_{m}(\tilde{\mathbf{R}});0\rangle=\frac{1}{\sqrt{N_{x}N_{y}}}\sqrt{\frac{1-e^{-2\alpha_{\mathbf{k}}\Delta_{z}}}{1-e^{-2\alpha_{\mathbf{k}}\Delta_{z}N_{z}}}}$
accounts for the weight of a localized exciton operator in a delocalized
one, such as the participation of $b_{\mathbf{n}_{0},0}^{\dagger}(0)$
in $b_{\mathbf{k}}^{\dagger}(0)$.
Eq. (\ref{coupling}) reveals that the maximal contribution of each double-polariton
Bogoliubov state to the energetic shift of the considered PES cross
section $E(R_{\mathbf{n}_{0},0},\mathbf{\tilde{0}}')$ is of the order
of $\frac{g_{\mathbf{k}}^{\mathbf{n}_{0},0}(0)}{\sqrt{N_{x}N_{y}}}$.
Considering macroscopic molecular ensembles with large $N\approx10^{7}$,
we computed Eq. \ref{second_order} by means of an integral approximation
over the polariton modes $\mathbf{k}$. 

\section{Results and discussion}
\subsection{Energetic effects}
\begin{figure}[h]
\includegraphics[scale=0.25]{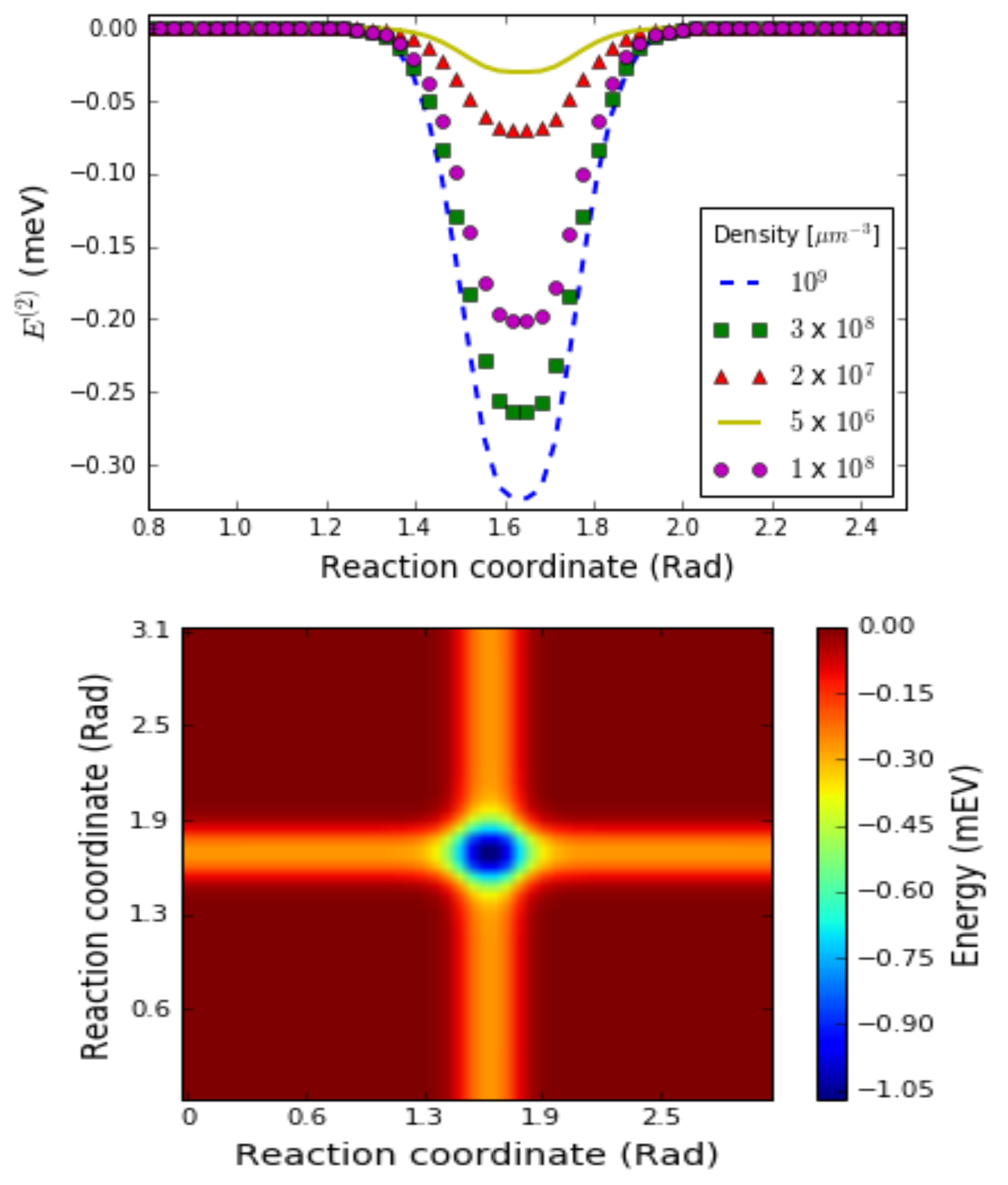} \caption{\emph{Upper:} Second order energy correction $E^{(2)}(R_{\mathbf{n}_{0},0},\mathbf{\tilde{0}}')$
of PES for one molecule isomerizing along the torsional coordinate
$R_{\mathbf{n}_{0},0}$; the rest of the molecules are fixed at the
equilibrium geometry. Calculations are displayed for various densities
$\rho$, keeping $W_{z}=120$ nm. Energy corrections are due to SP-exciton (see Eq. \ref{second_order}).
Note also that the energy scale of this correction is negligible in
comparison with the energy barrier of the reaction (see Fig. \ref{Adia_PESs}a).
\emph{Lower:} Same plot as in left, but for the 2D-ground-isomerization
PES of two molecules, keeping the configuration of the other molecules
at equilibrium ($E^{(2)}(R_{0},R_{1},\mathbf{\tilde{0}}')|$); the
density $\rho=3\times10^{8}\,\text{molecules}/\mu m^{3}$. \label{1D_PES}}
\end{figure}

\begin{figure}[b]
\includegraphics[scale=0.13]{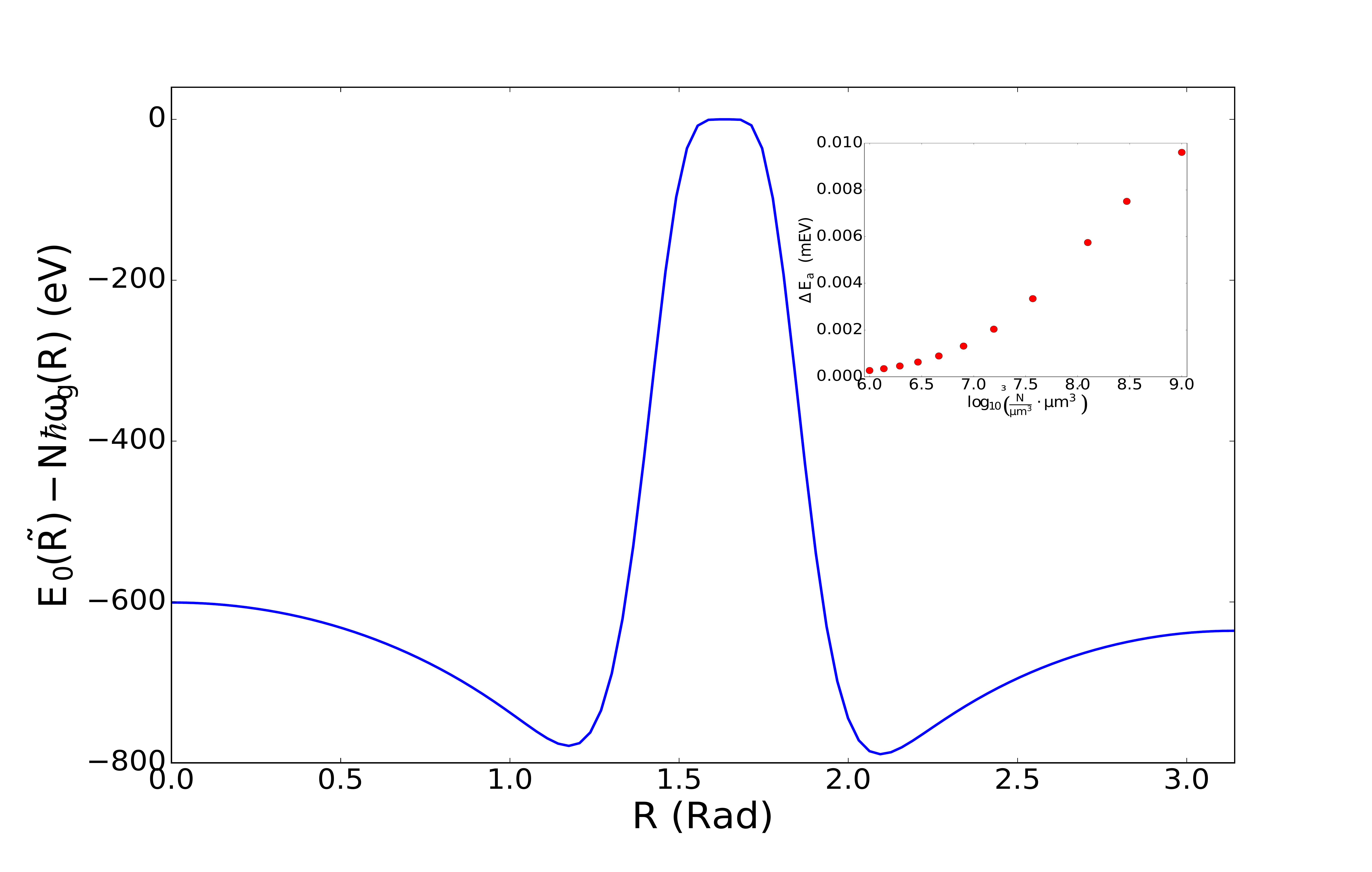} \caption{\emph{Main:} profile of the energy stabilization of the concerted
isomerization ($E_{0}(\tilde{\mathbf{R}})-N\hbar\omega_{g}(R)$, see
Eq. \ref{shift_vacuum}) of the whole molecular ensemble discussed
in the main text, due to the interaction with the plasmonic field.
We consider a molecular macroscopic ensemble ($N=8\times10^{7}$)
with density $\rho=3\times10^{8}\,\text{molecules}/\mu m^{3}$. \emph{Inset:}
molecular-density dependence of the the energy shift of the energy
barrier per molecule $|\Delta E_{a}|$ (see main text) due to the plasmonic
field, in this concerted scenario. \label{concerted}}
\end{figure}

We carry out our calculations with $\rho$ in the range of $10^{6}$
to $10^{9}$ molecules $\mu m^{-3}$ keeping $W_{z}=120\,\text{nm}$ (see Fig. \ref{1D_PES}); to obtain results in the thermodynamic limit, our calculations
take $N=8\times10^{7}$, even though the exact value is unimportant
as long as it is sufficiently large to give converged results. The
results displayed in Fig. \ref{1D_PES} show that the second order
energy corrections to the isomerization PES $E^{(2)}(R_{\mathbf{n}_{0},0},\mathbf{\tilde{0}}')$,
and in particular $E^{(2)}(R_{\mathbf{n}_{0},0}=R^{*},\mathbf{\tilde{0}}')\approx-0.25\,\text{meV}$,
are negligible in comparison with the bare activation barrier $E_{a}=\hbar\omega_{g}(R^{*})-\hbar\omega_{g}(0)=\hbar\omega_{g}(R^{*})\approx 1.8$ eV,
where $R^{*}\approx1.64$ rad corresponds to the transition
state. From Fig. \ref{Adia_PESs}b, we notice that there is a substantial
difference in SP-exciton coupling between the equilibrium ($R_{\mathbf{n}_{0},0}=0$)
and transition state geometries ($R_{\mathbf{n}_{0},0}=R^{*}$). Since
the perturbation in Eq. (\ref{perturbation}) is defined with respect
to the equilibrium geometry, $|E^{(2)}(R_{\mathbf{n}_{0},0},\mathbf{\tilde{0}}')|$
maximizes at the barrier geometry. To get some insight on the order
of magnitude of the result, we note that the sum shown in Eq. \ref{second_order}
can be very roughly approximated as
\begin{widetext}
\begin{align}
E^{(2)}(R_{\mathbf{n}_{0},0},\mathbf{\tilde{0}}')&=O\left[-\sum_{\mathbf{k}_{1}\leq\mathbf{k_{2}}}\frac{[g_{\mathbf{k}_{1}}^{\mathbf{n}_{0},0}(R_{\mathbf{n}_{0},0})]^{2}D_{\mathbf{k}_{2}}^{2}+[g_{\mathbf{k}_{2}}^{\mathbf{n}_{0},0}(R_{\mathbf{n}_{0},0})]^{2}D_{\mathbf{k}_{1}}^{2}}{(\hbar\omega_{\mathbf{k}_{1}}+\hbar\omega_{\mathbf{k}_{2}})/2+\hbar\omega_{e}(R_{\mathbf{n}_{0},0})}\right]\nonumber \\
 &=O\left[-\frac{1}{N_{x}N_{y}}\sum_{\mathbf{k}_{1}\leq\mathbf{k_{2}}}\frac{[g_{\mathbf{k}_{1}}^{\mathbf{n}_{0},0}(R_{\mathbf{n}_{0},0})]^{2}+[g_{\mathbf{k}_{2}}^{\mathbf{n}_{0},0}(R_{\mathbf{n}_{0},0})]^{2}}{(\hbar\omega_{\mathbf{k}_{1}}+\hbar\omega_{\mathbf{k}_{2}})/2+\hbar\omega_{e}(R_{\mathbf{n}_{0},0})}\right]\nonumber \\
 & =O\left[-\sum_{\mathbf{k}}\frac{[g_{\mathbf{k}}^{\mathbf{n}_{0},0}(R_{\mathbf{n}_{0},0})]^{2}}{\hbar\omega_{\mathbf{k}}+\hbar\omega_{e}(R_{\mathbf{n}_{0},0})}\right]\label{eq:lamb_shift_red}\\
 & =O\left(E_{LS}(R_{\mathbf{n}_{0},0})\right).\nonumber 
\end{align}
\end{widetext} In the first line, we used the fact that $\langle\mathbf{k}_{1},i;\mathbf{k}_{2},j|V(R_{\mathbf{n}_{0},0})|G(\tilde{\mathbf{0}})\rangle_{d}\approx [g_{\mathbf{k}_{1}}^{\mathbf{n}_{0},0}(R_{\mathbf{n}_{0},0})]^{2}D_{\mathbf{k}_{2}}^{2}+[g_{\mathbf{k}_{2}}^{\mathbf{n}_{0},0}(R_{\mathbf{n}_{0},0})]^{2}D_{\mathbf{k}_{1}}^{2}$
and averaged the Bogoliubov polariton excitation energies. In the
second line, assuming that the $\mathbf{k}\gg0$ values contribute
the most, we have $D_{\mathbf{k}}\approx\frac{1}{\sqrt{N_{x}N_{y}}}$.
Finally, in the third line, we have used the fact that the sum of
terms over $\mathbf{k}_{1},\mathbf{k}_{2}$ is roughly equal to $N_{x}N_{y}$
times a single sum over $\mathbf{k}$ of terms of the same order. The reason
why we are interested in the final approximation is because it corresponds
to the Lamb shift of a single isolated molecule, which can be calculated to
be $E_{LS}(0)=0.16\,\text{meV}$. Typically, Lamb
shift calculations require a cutoff to avoid unphysical divergences
\cite{Bethe1950}; we stress that in our plexciton model, this is
not necessary due to the decaying $|g_{\mathbf{k}}^{\mathbf{n}_{0},0}(R_{\mathbf{n}_{0},0})|$
as a function of $|\mathbf{k}|$. The fact that the corrections $E^{(2)}(R_{\mathbf{n}_{0},0},\mathbf{\tilde{0}}')$
have a similar order of magnitude to single-molecule Lamb shifts give
a pessimistic conclusion of harnessing USC to control ground-state
chemical reactions. 
Note, however, from calculations in Fig. \ref{1D_PES}, that there is variability in $E^{(2)}(R_{\mathbf{n}_{0},0},\mathbf{\tilde{0}}')$
as a function of molecular density (since density alters the character
of the Bogoliubov polaritons), although the resulting values are always
close to $E_{LS}(0)$. The molecular density cannot
increase without bound, since there exists a minimum molecular contact distance
determined by a van der Waals radius of the order of $0.3\,\text{nm}$ for organic molecules \cite{rowland1996intermolecular}, giving a maximum density
of $\rho\approx10^{10}\,\text{molecules}/\mu m^{3}$.\\
The results discussed so far describe the energy profile of the isomerization
of a single molecule keeping the rest at equilibrium geometry. It
is intriguing to inquire the effects of the SP field in a
concerted isomerization of two or more molecules, while keeping the
rest fixed at equilibrium geometry. Generalizing Eqs. (\ref{perturbation})\textendash (\ref{coupling})
to a two-molecule perturbation $V(R_{\mathbf{n}_{0},0},R_{\mathbf{n}_{1},0})$,
we computed the second order energetic corrections to the 2D-PES that
describe the isomerization of two neighbouring molecules at $\mathbf{n}_{0}$
and at $\mathbf{n}_{1}\equiv\mathbf{n}_{0}+\Delta_{x}\mathbf{\hat{x}}$,
keeping the other molecules fixed at $R_{\mathbf{n},s}=0$. The results
are reported in Fig. \ref{1D_PES} for $\rho=3\times10^{8}\,\text{molecules}/\mu m^{3}$, although outcomes of the same order of magnitude are obtained for the other densities considered in the one-dimensional case.
The two-dimensional PES cross-section $E^{(2)}(R_{\mathbf{n}_{0},0},R_{\mathbf{n}_{1},0},0,\cdots,0)\equiv E^{(2)}(R_{\mathbf{n}_{0},0},R_{\mathbf{n}_{1},0},\mathbf{\tilde{0}}')$
shows the existence of an energetic enhancement for the concerted
isomerization with respect to two independent isomerizations, \emph{i.e.}
$E^{(2)}(R_{\mathbf{n}_{0},0}=R^{*},R_{\mathbf{n}_{1},0}=R^{*},\mathbf{\tilde{0}}')\approx4E^{(2)}(R_{\mathbf{n}_{0},0}=R^{*},\mathbf{\tilde{0}}')$.
This enhacement is due to a constructive interference arising at the
amplitude level, $\langle\mathbf{k}_{1},i;\mathbf{k}_{2},j|V(R_{\mathbf{n}_{0},0}=R^{*},R_{\mathbf{n}_{1},0}=R^{*})|G(\tilde{\mathbf{0}})\rangle_{d}\approx2\langle\mathbf{k}_{1},i;\mathbf{k}_{2},j|V(R_{\mathbf{n}_{0},0}=R^{*})|G(\tilde{\mathbf{0}})\rangle_{d}$
for values of $\text{\ensuremath{\mathbf{k}}}_{1},\,\text{\ensuremath{\mathbf{k}}}_{2}\ll\frac{1}{\Delta_{x}}$,
such that the phase difference between the isomerizing molecules is negligible. Interestingly,
choosing the neighbouring molecules along the $x$ direction is important
for this argument; if instead we consider neighbours along $z$ (molecular
positions $\mathbf{n}_{0}$ and $\mathbf{n}_{0}+\Delta_{z}\mathbf{\hat{z}}$),
these interferences vanish and we approximately get the independent molecules result
$E^{(2)}(R_{\mathbf{n}_{0},0}=R^{*},R_{\mathbf{n}_{0},1}=R^{*},\mathbf{\tilde{0}}')\approx2E^{(2)}(R_{\mathbf{n}_{0},0}=R^{*},\mathbf{\tilde{0}}')$. \\
In light of the nontrivial energetic shift of the two-molecule case,
it is pedagogical to consider the SP effects on the cross-section
of the concerted isomerization of the whole ensemble,
even though it is highly unlikely that this kinetic pathway will be
of any relevance, especially considering the large barrier for the
isomerization of each molecule. Notice that the conservation of translational
symmetry in this scenario allows for the exact (nonperturbative) calculation
of the energetic shift $E_{0}(\tilde{\mathbf{R}})-N\hbar\omega_{g}(R)$
by means of Eq. \ref{shift_vacuum}. Our numerical calculations reveal
an energetic stabilization profile, which is displayed in Fig. \ref{concerted} for a molecular ensemble with $\rho=3\times 10^{8}$ molecules $\mu m^{-3}$.
As expected, we observe a stabilization of reactant and product regions
of the ground-state PES. This is a consequence of the transition dipole
moment being the strongest at those regions, as opposed to the transition
state, see Fig. \ref{Adia_PESs}b. However, even though these energetic
effects are of the order of hundreds of eV, they are negligible in
comparison with the total ground-state PES $N\hbar\omega_{g}(R)$,
or more specifically, to the transition barrier $NE_{a}=N\hbar\omega_{g}(R^{*})$
for the concerted reaction. 
Importantly, the change in activation energy per molecule in the concerted
isomerization with respect to the bare case $|\Delta E_{a}|=\left|\left(\frac{E_{0}(\tilde{\mathbf{R}}^{*})-E_{0}(\tilde{\mathbf{0}})}{N}-E_{a}\right)\right|\approx0.009\,\text{meV}$
is more than one order of magnitude smaller than the corresponding quantity $\left|E^{(2)}(R_{\mathbf{n}_{0},0}=R^{*},\mathbf{\tilde{0}}')\right|\approx0.25\,\text{meV}$
for the single-molecule isomerization case, see Fig. \ref{1D_PES}
and inset of Fig. \ref{concerted}. We believe that the reason for this trend is that
the isomerization of $n$ molecules, $n\ll N$, translates into a
perturbation which breaks the original translational symmetry of the
molecular ensemble. This symmetry breaking permits the interaction
of the molecular vacuum with the polaritonic $\mathbf{k}$-state reservoir
without a momentum-conservation restriction. This is reflected in
Eq. \ref{second_order}, where the sum is carried out over two not
necessarily equal momenta. In contrast, in the case of the concerted
isomerization of $N$ molecules, the  translational symmetry
of the system is preserved, which in turn restricts the coupling of the vacuum $|G(\tilde{\mathbf{0}})\rangle_{d}$
to excited states with $\mathbf{k}_{\mathrm{exc}}=-\mathbf{k}_{\mathrm{phot}}$.\\
Another intriguing observation is that, for this concerted isomerization,
the SP energetic effect per molecule $\frac{E_{0}(\tilde{\mathbf{R}})}{N}$
diminishes with the width of the slab $W_{z}$. This is the case given
that the SP quantization length $L_{\mathbf{k}}$ decays quickly with
$|\mathbf{k}|$ so that only the closest layers interact strongly
with the field. When we divide the total energetic effects due to
the SP modes by $N=N_{x}N_{y}N_{z}$, we obtain that $\frac{E_{0}(\tilde{\mathbf{R}})}{N}=O(\frac{1}{N_{z}})$
for large $W_{z}$.\\
The energetic shifts in all the scenarios discussed above are negligible
with respect to the corresponding energy barriers and the thermal
energy scale at room temperature which, unfortunately, signal the
irrelevance of USC to alter ground-state chemical reactivity for this
isomerization model. Although there is an overall (extensive) stabilization
of the molecular ensemble ground state, this effect is distributed across the ensemble, giving no possibility to alter the chemical reaction kinetics or thermodynamics considerably. However, we highlight the
intriguing interferences observed in the concerted isomerization processes.
Even though they will likely be irrelevant for this particular reaction,
they might be important when dealing with reactions with very low
barriers, especially when considering that these concerted pathways
are combinatorially more likely to occur than the single-molecule
events in the large $N$ limit. This is intriguing in light of the study carried out in \cite{galego2017many}, which discusses a different but related
effect of many reactions triggered by a single photon.\\
\subsection{Effects on non-adiabatic dynamics}
Finally, we discuss the importance of the nonadiabatic effects afforded
by nuclear kinetic energy. Previous works have considered the nonadiabatic
effects between polariton states at the level of SC \cite{Galego2015a,Kowalewski2016a}.
Alternatively, the consideration of nonadiabatic effects in USC for
a single molecule in a cavity was provided in \cite{Bennett2016a};
here, we address these issues for the many-molecule case and consider
both polariton and dark state manifolds. One could expect significantly
modified non-adiabatic dynamics about nuclear configurations where
the transition dipole moment magnitude $|\mathbf{\mathbf{\mu}}_{\mathbf{n},s}(R_{\mathbf{n},s})|$
is large, given a reduction in the energy gap between the ground and
the lower Bogoliubov polariton state. However, as we show below, this
energetic effect is not substantial due to the presence of dark states. \\
We consider the magnitude of the non-adiabatic couplings (NACs) for
the isomerization of a single molecule with reaction coordinate
$R_{\mathbf{n}_{0},0}$. For a region about $\tilde{\mathbf{R}}=\tilde{\mathbf{0}}$,
we estimate the magnitude of the NAC between $|G(\tilde{\mathbf{0}})\rangle_{d}$
and a state $|\mathbf{k},i\rangle=\xi_{\mathbf{k}}^{i\dagger}(0)|G(\tilde{\mathbf{0}})\rangle_{d}$
as: 
\begin{align}
|A_{\mathbf{k},i;g}(0)| & =\left|\langle\mathbf{k},i\big|\frac{\partial}{\partial R_{\mathbf{n}_{0},0}}\big|G(\mathbf{\tilde{0}})\rangle_{d}\right|\label{eq:NAC}\\
 & \approx\left|\beta_{\mathbf{k}}^{i}D_{\mathbf{k}}\left\langle e_{\mathbf{n}_{0},0}(0)\left|\frac{\partial}{\partial R_{\mathbf{n}_{0,0}}}\right|g_{\mathbf{n}_{0},0}(0)\right\rangle \right|,\nonumber 
\end{align}
where $|g_{\mathbf{n}_{0},0}(0)\rangle$($|e_{\mathbf{n}_{0},0}(0)\rangle$)
is the ground (excited) adiabatic state of the single molecule under
consideration (see Eq. (\ref{adiabatic_rep})) and we have ignored the
derivatives of $\beta_{\mathbf{k}}^{i}$ and $D_{\mathbf{k}}$ with
respect to $R_{\mathbf{n}_{0},0}$, assuming they are small at $\tilde{\mathbf{R}}=\tilde{\mathbf{0}}$,
where the chemical character of the Bogoliubov polariton states does
not change significantly with respect to nuclear coordinate. This is a consequence of the slowly changing transition dipole moment of the model molecule around $R_{\mathbf{n}_{0},0}=0$, see Fig. \ref{Adia_PESs}b.
Notice that we
have also assumed $\langle\mathbf{k},i|e_{\mathbf{n}_{0},0}(0)\rangle\approx\beta_{\mathbf{k}}^{i}D_{\mathbf{k}}$,
where we have used the fact that $\beta_{\mathbf{k}}^{i}\gg\gamma_{\mathbf{k}}^{i}$,
thus ignoring counterrotating terms, which as we have seen, give negligible
contributions. 
The time-evolution of a nuclear wavepacket in the ground-state will
be influenced by the Bogoliubov polariton states, each of which will
contribute with a finite probability of transition out of $\big|G(\mathbf{\tilde{0}})\rangle_{d}$.
From semiclassical arguments \cite{Bohm2003}, we can estimate the
transition probability $|C_{\mathbf{k}}^{i}(0)|^{2}$ for a nuclear
wavepacket on the ground-state PES at $\tilde{\mathbf{R}}=0$ to the
state $|\mathbf{k},i\rangle$, 
\begin{align}
|C_{\mathbf{k}}^{i}(0)|^{2} & \approx\left|\frac{\hbar v_{nuc}A_{\mathbf{k},i;g}(0)}{\hbar\omega_{i,\mathbf{k}}(0)-\hbar\omega_{g}(0)}\right|^{2}\label{eq:prob_NAC}\\
 & =\left|\frac{\hbar v_{nuc}\beta_{\mathbf{k}}^{i}D_{\mathbf{k}}}{\hbar\omega_{i,\mathbf{k}}(0)-\hbar\omega_{g}(0)}\right|^{2}\nonumber \\
 & \times\left|\left\langle e_{\mathbf{n}_{0},0}(0)\left|\frac{\partial}{\partial R_{\mathbf{n}_{0,0}}}\right|g_{\mathbf{n}_{0},0}(0)\right\rangle \right|^{2},\nonumber 
\end{align}
$v_{nuc}$ being the expectation value of the nuclear velocity. However, the Bogoliubov polariton $\mathbf{k}$-states
are only a small subset of the excited states of the problem. As mentioned
right after Eq. \ref{collective}, the plexciton setup contains
$N_{z}-1$ dark excitonic states for every $\mathbf{k}$ (eigenstates
of $H_{\mathrm{dark},\mathbf{k}}(0)$, see discussion
right after Eq. \ref{collective}); we ignore the very off-resonant couplings considered
in $H_{\mathrm{unklapp},\mathbf{k}}(0)$. The dark states also couple to $\big|G(\mathbf{\tilde{0}})\rangle_{d}$ non-adiabatically, with the corresponding transition
probability out of the ground state being, 
\begin{align}
|C_{\mathbf{k}}^{\mathrm{dark}}(0)|^{2} & \approx\sum_{Q}\left|\frac{\hbar v_{nuc}A_{\mathbf{k},Q;g}(0)}{\hbar\Delta(0)}\right|^{2}\label{eq:prob_NAC-1}\\
 & \approx P_{bare}(0)\left(\frac{1}{N_{x}N_{y}}-|D_{\mathbf{k}}|^{2}\right),\nonumber 
\end{align}
Here, we have summed over all dark states $Q$ for a given $\mathbf{k}$ and used $P_{bare}(0)=\left|\frac{v_{nuc}}{\Delta(0)}\right|^{2}\left|\left\langle e_{\mathbf{n}_{0},0}(0)\left|\frac{\partial}{\partial R_{\mathbf{n}_{0,0}}}\right|g_{\mathbf{n}_{0},0}(0)\right\rangle \right|^{2}$
to denote the probability of transition out of the ground state in
the absence of coupling to the SP field. In Eq. (\ref{eq:prob_NAC-1}) we used the fact that the
projection $|e_{\mathbf{n}_{0},0}(0)\rangle$ onto the dark $\mathbf{k}$
manifold of exciton states is $\left|\mathbf{P}_{\mathrm{dark},\mathbf{k}}(0)|e_{\mathbf{n}_{0},0}(0)\rangle\right|^{2}=\langle e_{\mathbf{n}_{0},0}(0)|\mathbf{I}_{\mathrm{exc},\mathbf{k}}(0)|e_{\mathbf{n}_{0},0}(0)\rangle-|D_{\mathbf{k}}|^{2}=\frac{1}{N_{x}N_{y}}-|D_{\mathbf{k}}|^{2}$,
with $\mathbf{P}_{\mathrm{dark},\mathbf{k}}(0)$ being the corresponding projector (see Eq. (\ref{dark_states})). 
We noticed that when $|\mathbf{k}|\to0$, the quantization volume $L_\mathbf{k}$ of the plasmonic field spans all the molecular-ensemble volume resulting in  completely delocalized bright and dark exciton
states across the different layers of the
slab, $\left|\mathbf{P}_{\mathrm{dark},\mathbf{k}}|e_{\mathbf{n}_{0},0}(0)\rangle\right|^{2}=\frac{N_{z}-1}{N}$,
and the dark states give the major contribution to the nonadiabatic
dynamics. On the other hand, when $|\mathbf{k}|\to\infty$, the plasmonic field interacts with the molecular layer at the bottom of the slab only and $\left|\mathbf{P}_{\mathrm{dark},\mathbf{k}}|e_{\mathbf{n}_{0},0}(0)\rangle\right|^{2} \to 0$.
The dark states do not participate, because the molecule located
at $\mathbf{n}_{0}$ only overlaps with the bright state which is
concentrated across the first layer of the slab (the dark states,
being orthogonal to the bright one, are distributed in the upper layers,
and do not overlap with $|e_{\mathbf{n}_{0},0}\rangle$).
With these results, we can compute the probability of transition out
of the ground-state $P_{out}$ as 
\begin{align}
P_{out}(0) & =\sum_{\mathbf{k}}\Bigg[\sum_{i}|C_{\mathbf{k}}^{i}(0)|^{2}+|C_{\mathbf{k}}^{\mathrm{dark}}(0)|^{2}\Bigg].\label{eq:P_out}
\end{align}
In view of the large off-resonant nature of most SP modes with respect to $\hbar\Delta(0)$ (see Fig. \ref{Collective-coupling}) and Eq. (\ref{eq:prob_NAC-1}), we have $\sum_{i}|C_{\mathbf{k}}^{i}(0)|^{2}\approx P_{bare}(0)|D_{\mathbf{k}}|^{2}$, such that $P_{out}(0)\approx P_{bare}(0)$. 
In our model, this is the
case, since the plexciton anticrossing occurs at small $|\mathbf{k}|$ and the SP energy quickly increases and reaches an asymptotic value after that point (see Fig. \ref{Collective-coupling}).\\
Using the parameters in \cite{Hoki2009a}, we obtain $\langle e(R_{\mathbf{n}_{0},0})|\frac{\partial}{\partial R_{\mathbf{n}_{0},0}}|g(R_{\mathbf{n}_{0},0})\rangle\approx0.01\,\text{\AA}^{-1}$, where we have assumed an effective radius of 1 $\text{\AA}$ for the isomerization mode of the model molecule.
We get an estimate of $v_{nuc}\approx 1\text{\AA}$ $\omega_{nuc}=1\text{\AA} \sqrt{\frac{k_{B}T}{m}}=9\times10^{10}\,\text{\AA} \text{s}^{-1}$
using $k_{B}=8.62\times10^{-5}\,\text{eV}\,\text{K}^{-1}$, $T=298\,\text{K}$ and
$m=2.5\,\text{amu}$ $\text{\AA}^{2}$. Finally, applying $\Delta(0)=3\,$ $\text{eV}$
gives $P_{bare}(0)\approx10^{-7}$, which is a negligible quantity.
A more pronounced polariton-effect is expected close to the PES avoided crossing. However, the rapid decay of the transition dipole
moment in this region (see Fig. \ref{Adia_PESs}a) precludes the formation of polaritonic states
that could have affected the corresponding nonadiabatic dynamics.
To summarize this part, even when the USC effects on the nonadiabatic dynamics
are negligible for our model, the previous discussion as well as Eq.
(\ref{eq:P_out}) distill the design principle that controls these
processes in other polariton systems: the plexciton anticrossings should happen
at large $\mathbf{k}$ values to preclude the overwhelming effects
of the dark states. This principle will be explored in future work
in other molecular systems.\\
The negligible polariton effect on the NACs, and the magnitude of
the energetic effects on the electronic energy landscape are strong
evidence to argue that the chemical yields and rates of the isomerization
problem in question remain intact with respect to the bare molecular
ensemble.

\section{Conclusions}

We showed in this work that, for the ground state landscape of a particular
isomerization model, there is no relevant collective stabilization
effect by USC to SPs which can significantly alter the kinetics or
thermodynamics of the reaction, in contrast with previous calculations
which show such possibilities in the Bogoliubov polariton landscapes
\cite{Galego2016,Herrera2016}. The negligible energetic corrections
to the ground-state PES per molecule can be approximated and interpreted as Lamb shifts
\cite{Bethe1950} experienced by the molecular states due to the interaction
with off-resonant plasmonic modes. The key dimensionless parameter which determines
the USC effect on the ground-state PES is the ratio of the individual
coupling to the transition frequency $g_{\mathbf{k}}^{\mathbf{n},s}/\hbar\Delta$.
This finding is similar to the conclusions of a recent work \cite{Galego2015a,Cwik2016}.
In particular, it is shown in \cite{Cwik2016} that the rotational
and vibrational degrees of freedom of molecules exhibit a self-adaptation
which only depends on light-matter coupling at the single-molecule
level. Therefore, more remarkable effects are expected in the regime
of USC of a single molecule interacting with an electric field. To
date, the largest single molecule interaction energy achievable experimentally
is around 90 meV \cite{Chikkaraddy2016} in an ultralow nanostructure
volume. This coupling strength is almost two orders of magnitude larger
that those in our model. Also, previous works have shown \cite{Niemczyk2010,Jenkins2013}
that this regime is achievable for systems with transition frequencies
on the microwave range. Additionally, the experimental realization of vibrational USC has been carried out
recently \cite{george2016multiple}. The latter also suggests the theoretical exploration of USC
effects on chemical reactivity at the rotational or vibrational energy scales, where the
energy spacing between levels is significantly lower than typical electronic
energy gaps. \\
We highlighted some intriguing quantum-coherent effects where concerted
reactions can feature energetic effects that are not incoherent combinations of the bare molecular
processes. These interference effects are unlikely to play an important
role in reactions exhibiting high barriers compared to $k_{B}T$.
However, they might be important for low-barrier processes, where
the number of concerted reaction pathways becomes combinatorially
more likely than single molecule processes. On the other hand, we
also established that, due to the large number of dark states in these
many-molecule polariton systems, nonadiabatic effects are not modified
in any meaningful way under USC, at least for the model system explored.
We provided a rationale behind this conclusion and discussed possibilities
of seeing modifications in other systems where the excitonic and the
electromagnetic modes anticross at large $\mathbf{k}$ values.

Finally, it is worth noting that even though we considered an ultrastrong
coupling regime ( $\sqrt{\mathcal{N}_{\mathbf{k}}(R)}$ reaches more
than $10\%$ of the maximum electronic energy gap in our model \cite{Moroz}),
the system does not reach a Quantum Phase Transition (QPT) \cite{Emary,Li2006a}.
In our model, this regime would require high density ($\sim 10^{10}\,\text{molecules}$
$\mu m^{-3}$) samples, keeping $\mu\approx 2 $ $e\text{\AA}$. The implications
of this QPT on chemical reactivity have not been explored in this
work, but are currently being studied in our group. To conclude, our present work
highlights the limitations but also possibilities of USC in the context
of control of chemical reactions using polaritonic systems.\\

\section{Acknowledgments}

R.F.R., J.C.A., and J.Y.Z. acknowledge support from the NSF CAREER award
CHE-1654732. L.A.M.M is grateful to the support of the UC-Mexus CONACyT
scholarship for doctoral studies. All authors acknowledge generous
startup funds from UCSD. L.A.M.M. and J.Y.Z are thankful to Prof. Felipe Herrera for useful discussions.

\section{Appendix}

In this appendix we outline the perturbative methodology that leads
to the equations shown in the main text. Under the perturbative approach,
it is convenient to express the perturbation in Eq. (\ref{perturbation})
in terms of the Bogoliubov operators defined by Eq. (\ref{eq:bogoliubov_quasi}).
Notice that Eq. \ref{perturbation} introduces $R$-dependent exciton
operators, while the zeroth order eigenstates (the polariton quasiparticles)
are defined for all molecules at the configuration $R_{\mathbf{n},s}=0$.
It would be useful to find a relation $b_{\mathbf{n}_{0},0}^{\dagger}(R)=f(b_{\mathbf{n}_{0},0}^{\dagger}(0),b_{\mathbf{n}_{0},0}(0))$
for any $R$, in order to carry out the aforementioned change of basis. 

The function $f(b_{\mathbf{n}_{0},0}^{\dagger}(0),b_{\mathbf{n}_{0},0}(0))$
can be found by working on the diabatic basis (see Eq. \ref{adiabatic_rep}).
For any operator $b_{\mathbf{n},s}^{\dagger}(R)$, using $b_{\mathbf{n},s}^{\dagger}(R)=|e_{\mathbf{n},s}(R)\rangle\langle g_{\mathbf{n},s}(R)|$,
we have
\begin{align}
b_{\mathbf{n},s}^{\dagger}(R) & =\left[-\sin(\theta(R)/2)|\text{trans}_{\mathbf{n},s}\rangle+\cos(\theta(R)/2)|\text{cis}_{\mathbf{n},s}\rangle\right]\nonumber \\
 & \times\left[\cos(\theta(R)/2)\langle\text{trans}_{\mathbf{n},s}|+\sin(\theta(R)/2)\langle\text{cis}_{\mathbf{n},s}|\right]\nonumber \\
 & =-\sin(\theta(R)/2)\cos(\theta(R)/2)b_{\mathbf{n},s}(0)b_{\mathbf{n},s}^{\dagger}(0)\label{exc_adia}\\
 & -\sin(\theta(R)/2)^{2}b_{\mathbf{n},s}(0)+\cos(\theta(R)/2)^{2}b_{\mathbf{n},s}^{\dagger}(0)\nonumber \\
 & +\cos(\theta(R)/2)\sin(\theta(R)/2)b_{\mathbf{n},s}^{\dagger}(0)b_{\mathbf{n},s}(0),\nonumber 
\end{align}
where we used $b_{\mathbf{n},s}^{\dagger}(R_{eq})=|\text{cis}_{\mathbf{n},s}\rangle\langle\text{trans}_{\mathbf{n},s}|$.
In light of Eq. \ref{exc_adia} we notice that the perturbation (\ref{perturbation})
in the second row produces chains with up to four exciton operators.
In view of the delocalized nature of the zeroth-order eigenstates
and the localized character of the exciton operators $b_{\mathbf{n},s}(0)$,
we have that the matrix elements that appear in $E^{(2)}(R_{\mathbf{n}_{0},0},\mathbf{\tilde{0}}')$
are of the form $\langle G(\tilde{\mathbf{0}})|_{d}\xi_{\mathbf{k}_{1}}^{i}\xi_{\mathbf{k}_{2}}^{j}\dots\xi_{\mathbf{k}_{m+1}}^{l}F_{m}(b_{\mathbf{n}_{0},0}^{\dagger}b_{\mathbf{n}_{0},0})Z(a_{\mathbf{k}}^{\dagger},a_{\mathbf{k}})|G(\tilde{\mathbf{0}})\rangle_{d}\approx O(1/(N_{x}N_{y})^{m/2})$,
where $F_{m}(b_{\mathbf{n}_{0},0}^{\dagger}b_{\mathbf{n}_{0},0})$
stands for a chain with $1\leq m\leq4$ exciton operators and $Z(a_{\mathbf{k}}^{\dagger},a_{\mathbf{k}})$
is a function of a single photonic operator. In the macroscopic limit
$1\ll N_{x}N_{y}$, we can neglect chains $F_{m}(b_{\mathbf{n}_{0},0}^{\dagger}b_{\mathbf{n}_{0},0})Z(a_{\mathbf{k}}^{\dagger},a_{\mathbf{k}})$
for $m\geq2$. This leads to the simplification,
\begin{align}
b_{\mathbf{n}_{0},0}^{\dagger}(R)+b_{\mathbf{n}_{0},0}(R) & \approx\left(\cos(\theta(R)/2)^{2}-\sin(\theta(R)/2)^{2}\right)\label{eq:approx_exc}\\
 & \times\left(b_{\mathbf{n}_{0},0}^{\dagger}(0)+b_{\mathbf{n}_{0},0}(0)\right),\nonumber 
\end{align}
and the perturbation acquires the simple form,
\begin{align}
V(R) & \approx\sum_{\mathbf{k},\mathbf{k}_{1}}D_{\mathbf{k}_{1}}^{\mathbf{n}_{0},0}\left(\cos(\theta(R))g_{\mathbf{k}}^{\mathbf{n}_{0},0}(R)-g_{\mathbf{k}}^{\mathbf{n}_{0},0}(0)\right)\nonumber \\
 & \times\left(b_{\mathbf{k}_{1}}(0)a_{\mathbf{k}}^{\dagger}+b_{\mathbf{k}_{1}}^{\dagger}(0)a_{\mathbf{k}}^{\dagger}+a_{\mathbf{k}}b_{\mathbf{k}_{1}}(0)+b_{\mathbf{k}_{1}}^{\dagger}(0)a_{\mathbf{k}}\right).\label{eq:approx_pert}
\end{align}
To write this last expression in terms of the Bogoliubov operators
$\{\xi_{\mathbf{k_{1}}}^{i\dagger}(0)\xi_{\mathbf{k}_{2}}^{j\dagger}(0)\}$
we start from the transformation $\vec{\xi}=T\vec{b}$ : 
\begin{equation}
\begin{bmatrix}\xi_{\mathbf{k}}^{L}(0)\\
\xi_{\mathbf{k}}^{U}(0)\\
\xi_{\mathbf{-k}}^{L\dagger}(0)\\
\xi_{\mathbf{-k}}^{U\dagger}(0)
\end{bmatrix}=\begin{bmatrix}\alpha_{\mathbf{k}}^{L} & \beta_{\mathbf{k}}^{L} & \gamma_{\mathbf{k}}^{L} & \delta_{\mathbf{k}}^{L}\\
\alpha_{\mathbf{k}}^{U} & \beta_{\mathbf{k}}^{U} & \gamma_{\mathbf{k}}^{U} & \delta_{\mathbf{k}}^{U}\\
\gamma_{\mathbf{k}}^{L*} & \delta_{\mathbf{k}}^{L*} & \alpha_{\mathbf{k}}^{L*} & \beta_{\mathbf{k}}^{L*}\\
\gamma_{\mathbf{k}}^{U*} & \delta_{\mathbf{k}}^{U*} & \alpha_{\mathbf{k}}^{U*} & \beta_{\mathbf{k}}^{U*}
\end{bmatrix}\begin{bmatrix}a_{\mathbf{k}}\\
b_{\mathbf{k}}(0)\\
a_{\mathbf{-k}}^{\dagger}\\
b_{-\mathbf{k}}^{\dagger}(0)
\end{bmatrix}.\label{matrix_bogo}
\end{equation}
From the matrix representation of the normalization $|\alpha_{\mathbf{k}}^{i}|^{2}+|\beta_{\mathbf{k}}^{i}|^{2}-|\gamma_{\mathbf{k}}^{i}|^{2}-|\delta_{\mathbf{k}}^{i}|^{2}=1$ \cite{Ciuti2005},
it follows that, 
\begin{equation}
TI_{-}T^{\dagger}=I_{-},\label{matrix_norm}
\end{equation}
where 
\begin{equation}
I_{-}=\begin{bmatrix}1 & 0 & 0 & 0\\
0 & 1 & 0 & 0\\
0 & 0 & -1 & 0\\
0 & 0 & 0 & -1
\end{bmatrix}.\label{norm}
\end{equation}
We also have that $T^{-1}=I_{-}T^{\dagger}I_{-}$ and that
\begin{equation}
T^{-1}=\begin{bmatrix}\alpha_{\mathbf{k}}^{L*} & \alpha_{\mathbf{k}}^{U*} & -\gamma_{\mathbf{k}}^{L} & -\gamma_{\mathbf{k}}^{U}\\
\beta_{\mathbf{k}}^{L*} & \beta_{\mathbf{k}}^{U*} & -\delta_{\mathbf{k}}^{L} & -\delta_{\mathbf{k}}^{U}\\
-\gamma_{\mathbf{k}}^{L*} & -\gamma_{\mathbf{k}}^{U*} & \alpha_{\mathbf{k}}^{L} & \alpha_{\mathbf{k}}^{U}\\
-\delta_{\mathbf{k}}^{L*} & -\delta_{\mathbf{k}}^{U*} & \beta_{\mathbf{k}}^{L} & \beta_{\mathbf{k}}^{U}
\end{bmatrix}.\label{eq:T-1}
\end{equation}
Using Eq. \ref{eq:T-1}, we can readily evaluate $\vec{b}=T^{-1}\vec{\xi}$.
From this relationship, the change of the localized operators to the
Bogoliubov basis is accomplished. Finally, the matrix elements to
compute $E^{(2)}(R_{\mathbf{n}_{0},0},\mathbf{\tilde{0}}')$ can be
evaluated by means of Wick's theorem,
\begin{align}
\langle G(\tilde{\mathbf{0}})|_{d}\xi_{\mathbf{k}_{1}}^{l}\xi_{\mathbf{k}_{2}}^{n}\xi_{\mathbf{k}}^{i\dagger}\xi_{\mathbf{k}'}^{j\dagger}|G(\tilde{\mathbf{0}})\rangle_{d} & =\delta_{l,j}\delta_{\mathbf{k}_{1},\mathbf{k}'}\delta_{i,n}\delta_{\mathbf{k}_{2},\mathbf{k}}\label{wicks}\\
 & +\delta_{n,j}\delta_{\mathbf{k}_{2},\mathbf{k'}}\delta_{m,i}\delta_{\mathbf{k}_{1},\mathbf{k}},\nonumber 
\end{align}
leading to Eq. \ref{coupling}.

\bibliographystyle{achemso}
\bibliography{usc}

\end{document}